\documentclass{article}
\usepackage{spconf,amsmath,graphicx,hyperref}

\usepackage{cite}
\usepackage{amssymb,amsfonts}
\usepackage{textcomp}
\usepackage{xcolor}
\usepackage{booktabs}
\usepackage{multirow}
\usepackage{subcaption}
\hypersetup{hidelinks}

\newcommand{\citeme}[1]{{\color{blue}[C]}}

\def\BibTeX{{\rm B\kern-.05em{\sc i\kern-.025em b}\kern-.08em
    T\kern-.1667em\lower.7ex\hbox{E}\kern-.125emX}}

\title{Synth-JEPA: Joint Embedding Prediction for Renderer-Free Synthesizer Parameter Search}
\name{Ben Hayes$^1$, Haokun Tian$^{1,2}$, Stefan Lattner$^1$
}
\address{$^1$Sony Computer Science Laboratories, Paris, France\\
$^2$Queen Mary University of London, UK}
\begin{document}
%
\maketitle
\begin{abstract}
Sound matching can be formulated as optimizing synthesizer parameters against an audio-domain objective.
However, objectives derived from generic audio representations are often difficult to optimize, while direct search requires rendering every candidate.
We introduce \emph{Synth-JEPA}, which learns mutually predictive audio and parameter representations from paired synthesizer data.
At inference, candidate parameters are scored directly in this learned space, yielding a renderer-free objective whose audio geometry is shaped by parameter correspondences rather than generic audio similarity.
We evaluate Synth-JEPA on Surge XT using held-out synthesizer sounds and out-of-domain NSynth and FSD50K targets, against inverse models, direct search, and learned proxy objectives.
Synth-JEPA outperforms all baselines in-domain and remains competitive out-of-domain.
Its matching quality continues to improve with additional test-time search, allowing compute to be traded for match quality. In pairwise listening tests, listeners preferred Synth-JEPA in 85\% of trials overall.
Together, these results show that an audio representation with a parameter-induced geometry allows synthesizer sound matching to be approached as an effective renderer-free search problem.
\end{abstract}
\section{Introduction}
\begin{figure}[t]
    \centering
    \includegraphics[width=\columnwidth]{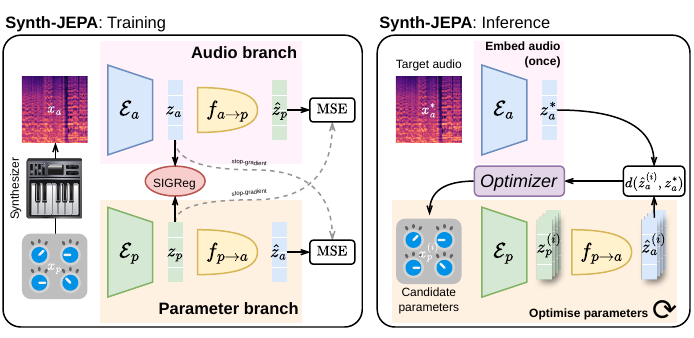}
    \caption{\textbf{Synth-JEPA training and inference.} $\mathcal{E}_a$ and $\mathcal{E}_p$ are audio and parameter encoders, while $f_{a\rightarrow p}$ and $f_{p\rightarrow a}$ are cross-domain predictors. At inference, target audio is encoded once and candidate parameters are optimized directly against the learned objective, without rendering candidate audio during search.}
    \label{fig:synth-jepa}
\end{figure}
Modern audio synthesizers are intricate systems, combining many methods for sound production and manipulation with large numbers of interacting controls.
Selecting parameters to approximate a target sound is therefore challenging. Further, many synthesizers can produce similar or equivalent signals from different parameter configurations, meaning the corresponding inverse mapping is generally ill-posed~\cite{hayes_audio_2025}.

Existing approaches broadly either optimize synthesizer parameters separately for each target~\cite{horner_machine_1993a,yee-king_automatic_2018,shier_spiegelib_2020}, or learn an amortized mapping from audio to parameters using supervised estimation~\cite{bruford_synthesizer_2024}, conditional generative models~\cite{esling_flow_2020,vaillant_improving_2021,hayes_audio_2025}, differentiable synthesis~\cite{masuda_synthesizer_2021,yang_white_2023}, or reward-based training~\cite{shin_synthrl_2025}.
Search methods evaluate candidate parameters according to an audio-domain objective, but typically require rendering each candidate with the synthesizer.
Learned proxies can reduce this cost by predicting audio representations from parameters~\cite{barkan_inversynth_2023,combes_neural_2025}, but search quality then depends both on the accuracy of the proxy and on the choice of representation.
In particular, general audio representations are not learned from the relationship between synthesizer parameters and the sounds they produce.

We propose \emph{Synth-JEPA}, a joint audio--parameter predictive model used as an objective for parameter search, which learns mutually predictive representations of paired synthesizer parameters and rendered audio.
This makes the learned audio representation depend directly on parameter--audio correspondences. At inference, the target audio is encoded once and candidate parameters are scored using the learned parameter encoder and cross-domain predictor.
Search can therefore be conducted directly in the audio embedding space, without any calls to the synthesizer.

Most closely related to our representation-learning setup, Braun and Finkelstein adapt the SLAP EMA-teacher objective~\cite{guinot_slap_2025} to learn joint DX7 audio--parameter embeddings for preset retrieval~\cite{braun_fm_2026}.
Here, we instead use a joint audio--parameter representation as the objective for iterative parameter search, and compare the SLAP-style EMA objective with SIGReg~\cite{balestriero_lejepa_2025} within the same Synth-JEPA framework.

Audio examples are available online.~\footnote{\url{https://benhayes.net/synth-jepa}}



\section{Synth-JEPA}
\label{sec:method}

Let $\mathcal{P}$ denote the synthesizer parameter space, which contains both continuous and discrete controls, and let $\mathcal{S}\subset\mathbb{R}^{n}$ denote the space of audio signals.
We write the synthesizer as $S:\mathcal{P}\rightarrow\mathcal{S}$.
Our training data consists of pairs $(x, y)$, where $x\in\mathcal{P}$ is sampled from the training parameter distribution and $y=S(x)$ is the corresponding rendered signal.
We sample parameters from a uniform prior and, following prior work~\cite{hayes_audio_2025}, map continuous parameters to $[-1, 1]$ and adopt a one-hot encoding for discrete parameters.
The sampled parameter distribution induces a generally non-uniform distribution over $\mathcal{S}$, which may therefore affect the geometry learned for search.
Future work will explore audio-aware sampling strategies.

\subsection{Cross-domain predictive learning}
Audio and parameters are mapped to separate representations,
\begin{equation}
    z_a = \mathcal{E}_a(y), \qquad z_p = \mathcal{E}_p(x),
\end{equation}

\noindent
with audio and parameter encoders $\mathcal{E}_a$ and $\mathcal{E}_p$, respectively. These are then mapped to the opposite modality using cross-domain predictors,

\begin{equation}
    \hat z_p=f_{a\rightarrow p}(z_a), \qquad
    \hat z_a=f_{p\rightarrow a}(z_p).
\end{equation}

\noindent
We train these mappings with the predictive objective
\begin{equation}
\mathcal{L}_{\mathrm{pred}}
= \operatorname{MSE}(\hat z_p,\operatorname{sg}(z_p))
+ \operatorname{MSE}(\hat z_a,\operatorname{sg}(z_a)),
\label{eq:pred}
\end{equation}
where $\operatorname{sg}$ is the stop-gradient operator. This setup is illustrated in Fig.~\ref{fig:synth-jepa}.

To prevent collapse we use SIGReg, the regularization strategy proposed in LeJEPA~\cite{balestriero_lejepa_2025}, to regularize the encoders' output distributions towards an isotropic Gaussian. 
We apply SIGReg to each encoder branch independently, following the setup in LeVLJEPA~\cite{kuhn_levljepa_2026}.
To explore the sensitivity of the resulting embedding to the choice of anti-collapse mechanism, we also train an \emph{EMA teacher} variant using the joint-embedding objective proposed in SLAP \cite{guinot_slap_2025} which is used recently for FM parameter-audio joint representation learning~\cite{braun_fm_2026}.

\subsection{Renderer-free parameter search}\label{sec:search}
Given target audio $y^\star$, we compute $z_a^\star=\mathcal{E}_a(y^\star)$ once. Candidate parameters are then scored by predicting their representation in the audio space,
\begin{equation}
    D_{\mathrm{JEPA}}(y^\star,x)
    = \mathrm{MSE}\!\left(z_a^\star,
    f_{p\rightarrow a}(\mathcal{E}_p(x))\right),
\end{equation}

\noindent which, importantly, doesn't require a synthesizer call.
We thus estimate

\begin{equation}
    \hat x
    = \arg\min_{x\in\mathcal{P}}
    D_{\mathrm{JEPA}}(y^\star,x).
    \label{eq:search}
\end{equation}

Since $D_\mathrm{JEPA}$ is differentiable, Eq.~\ref{eq:search} can be solved with gradient-free or gradient-based optimizers.
We adopt a hybrid two-stage approach.
First, we spend half the evaluation budget evolving a population of 32 candidates with
JADE~\cite{zhang_jade_2009}.
Mutation and crossover operators are applied to continuous parameters, while categorical parameters are either copied from another candidate or resampled.
Second, the eight best candidates as scored by $D_\mathrm{JEPA}$ are further refined by gradient descent on their continuous parameters using Adam with an initial learning rate of $0.1$ and cosine decay to zero.
During this refinement stage, we prune the candidate pool three times at evenly spaced intervals, each time discarding the worse-performing half according to \(D_\mathrm{JEPA}\). After the final pruning step, the sole remaining candidate is returned.


\section{Experiments}
\label{sec:experiments}

\begin{table*}[ht]
\centering
\caption{
Sound matching results for in-domain Surge XT presets drawn from the parameter prior, and for out-of-domain NSynth and FSD50K targets, 1\,024 targets per set. Bold and underlining mark the best and second-best result in each column.
}
\label{tab:main}
\scriptsize
\setlength{\tabcolsep}{2.4pt}
\renewcommand{\arraystretch}{1.15}
\resizebox{\textwidth}{!}{%
\begin{tabular}{ll ccc ccc ccc cc}
\toprule
&&
\multicolumn{3}{c}{\textbf{In-domain presets}} &
\multicolumn{3}{c}{\textbf{NSynth}} &
\multicolumn{3}{c}{\textbf{Freesound (FSD50K)}} & \\
\cmidrule(lr){3-5}
\cmidrule(lr){6-8}
\cmidrule(lr){9-11}

\textbf{Family} &
\textbf{Method} &
MSS $\downarrow$ & wMFCC $\downarrow$ & CLAP $\uparrow$ &
MSS $\downarrow$ & wMFCC $\downarrow$ & CLAP $\uparrow$ &
MSS $\downarrow$ & wMFCC $\downarrow$ & CLAP $\uparrow$ &
\textbf{Renders / target} &
\textbf{Parameters} \\
\midrule

\multirow{2}{*}{Synth-JEPA}
& SIGReg
& \textbf{8.44} & \textbf{11.07} & \textbf{0.793} & \textbf{10.25} & \underline{15.54} & \textbf{0.465} & \underline{13.59} & 14.81 & \underline{0.250}
& 0 & 53.3M \\
& EMA Teacher
& 11.52 & 15.29 & 0.625 & 14.02 & 19.33 & 0.323 & 15.39 & 17.33 & 0.179
& 0 & 53.3M \\
\midrule
\multirow{2}{*}{Amortized}
& AST Regression
& 17.94 & 23.16 & 0.499 & 28.06 & 30.64 & 0.104 & 21.48 & 26.56 & 0.086
& 0 & 25.0M \\
& Flow Matching
& 12.75 & 14.78 & 0.707 & 16.67 & 21.24 & 0.356 & 17.94 & 18.84 & 0.204
& 0 & 44.0M \\
\midrule
\multirow{3}{*}{Renderer search}
& Log-Mel distance
& 15.86 & \underline{12.17} & 0.553 & 17.81 & \textbf{14.10} & 0.199 & 14.55 & \textbf{11.61} & 0.178
& 2048 & --- \\
& Embedding distance (mn20)
& 12.90 & 16.43 & 0.730 & 14.61 & 25.05 & 0.387 & 16.12 & 21.41 & 0.249
& 2048 & 17.9M \\
& Learned $\mathcal{E}_a$ distance
& \underline{9.48} & 13.15 & 0.729 & 12.31 & 17.30 & 0.360 & 14.49 & 16.88 & 0.208
& 2048 & 17.9M \\
\midrule
\multirow{2}{*}{Proxy search}
& Embedding (mn20) proxy
& 12.56 & 15.77 & \underline{0.768} & 13.65 & 23.98 & \underline{0.436} & 15.44 & 19.79 & \textbf{0.261}
& 0 & 31.8M \\
& Audio (log-Mel) proxy
& 10.75 & 12.74 & 0.699 & \underline{11.03} & \textbf{14.10} & 0.350 & \textbf{12.18} & \underline{13.87} & 0.216
& 0 & 31.8M \\

\bottomrule
\end{tabular}%
}
\vspace{-2mm}
\end{table*}

\subsection{Data and tasks}
Following prior work~\cite{hayes_audio_2025}, we generate training data using the Surge XT software synthesizer.
We adopt a set of 139 active parameters, similar to those selected in [11], but disable audio effects.
We disable any modulator settings known to introduce non-determinism (e.g. Sample \& Hold).
Rather than fixing a dataset size, we render in-domain audio online during training.
Audio is rendered at 44.1 kHz in stereo with a duration of 3.0 seconds. 
Audio encoder inputs are log-Mel spectrograms with a 25 ms window and 10 ms hop.
To avoid hand-tuned input normalization, we estimate channel-wise mean and variance over the first 8k training spectrograms using Welford’s online algorithm~\cite{welford_note_1962} and freeze the resulting statistics for the remainder of training.

We first evaluate on held-out \emph{in-domain} targets generated by the same synthesizer and rendering protocol. 
We then evaluate general sound matching on the test splits of NSynth~\cite{engel_neural_2017} and FSD50K~\cite{fonseca_fsd50k_2022}, trimmed to the first 3s of each clip, resampled to 44.1 kHz, and upmixed to stereo.

\subsection{Models and baselines}
Synth-JEPA's audio encoder $\mathcal{E}_a$ is a transformer operating on stereo log-Mel spectrograms (128 bands, 25\,ms window, 10\,ms hop), projected via 1D convolution to 150 tokens.
A learned summary token is concatenated to the sequence and its output is taken as $z_a\in\mathbb{R}^{512}$.
The parameter encoder $\mathcal{E}_p$ represents each parameter with a linear projection plus a parameter-specific bias. 
A Perceiver bottleneck~\cite{jaegle_perceiver_2021} with 32 learned latents cross-attends to these tokens, followed by eight self-attention blocks.
Averaging the final outputs gives $z_p\in\mathbb{R}^{512}$.
Both encoders use an internal dimension of 512, with 8 attention heads.
The predictors $f_{p\rightarrow a}$ and $f_{a\rightarrow p}$ are three-block residual MLPs with hidden width 1024 and linear output heads.
The full model has 53M parameters. 
To understand the effect of the SIGReg anti-collapse term, we also train a version of Synth-JEPA which instead uses an EMA teacher model~\cite{braun_fm_2026,guinot_slap_2025}.
We train for 1M steps with batch size 64, using AdamW with a learning rate of $3\times10^{-4}$ with a warmup-stable-decay scheduler~\cite{wen_understanding_2025}, and a weight decay of $0.05$.

We compare against two amortized predictive baselines.
The first performs regression with an audio spectrogram transformer~\cite{bruford_synthesizer_2024}, while the second uses a symmetry-aware parameter domain flow matching model~\cite{hayes_audio_2025}. 
We adopt the same training recipe for both.

We also compare to direct, or ``renderer-in-the-loop'' (RITL) search, in which each candidate is rendered by the synthesizer.
For this setup, we score candidates using (i) log-Mel $L_1$ distance; (ii) a pretrained audio embedding (we adopt the mn20 variant of EfficientAT~\cite{schmid_efficient_2023}, following \cite{combes_neural_2025}); or (iii) our trained Synth-JEPA audio encoder $\mathcal{E}_a$. 

To disentangle the effect of amortization from that of parameter-induced embedding geometry, we also train two neural proxies which predict audio representations from parameter input.
Like the RITL baselines, the first directly predicts a log-Mel spectrogram of the synthesizer's audio.
The second follows the work of Combes et al.~\cite{combes_neural_2025} and predicts a pretrained audio embedding (EfficientAT mn20~\cite{schmid_efficient_2023}).

As the synthesizer is not differentiable, all RITL variants are restricted to using only JADE~\cite{zhang_jade_2009} for their full search budget, while Synth-JEPA and the proxy objectives additionally permit gradient-based refinement (Sec.~\ref{sec:search}).
The $\mathcal{E}_a$ RITL variant thus helps isolate the effect of embedding geometry from any benefit from gradient descent. 

\vspace{-2mm}
\subsection{Evaluation}
All metrics are computed after rendering the returned parameters. We report multi-scale spectral distance (MSS), warped MFCC distance (wMFCC), following~\cite{hayes_audio_2025}, and cosine similarity between fixed CLAP~\cite{wu_large-scale_2023} embeddings of the target and reconstruction.
Wall-clock times are measured per target on an RTX 3090.

\vspace{-2mm}
\section{Results}
\label{sec:results}

\begin{figure}[t]
\centering
\includegraphics[width=\columnwidth]{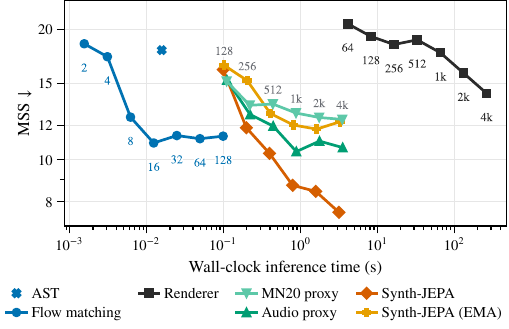}
\caption{Sound matching performance when scaling the number of model evaluations (search budget / ODE solver steps). Wall-clock time reported per-target on one RTX 3090. Labels indicate number of model/objective evaluations per target.}
\label{fig:compute-methods}
\end{figure}

\subsection{Sound-matching performance}

Table~\ref{tab:main} shows sound matching results for all methods, with a budget of 2048 objective evaluations for search-based methods, and 20 ODE steps for flow matching.
In-domain, Synth-JEPA (SIGReg) obtains the best result across all metrics.
Its performance remains strong outside the training distribution, achieving best or second-best performance on all metrics except wMFCC, where log-Mel-based objectives lead.

The learned $\mathcal{E}_a$ representation gives the strongest renderer-in-the-loop MSS results across all datasets, indicating that the parameter-induced embedding geometry itself benefits search.
Synth-JEPA (SIGReg) also broadly outperforms both proxy baselines in-domain and on NSynth, though its performance is more mixed on Freesound's wider audio distribution.

SIGReg outperforms the EMA-teacher variant on every dataset and metric.
This suggests that the choice of anti-collapse mechanism exerts a meaningful influence on the resulting optimization geometry.

\subsection{Inference-time scaling}

\begin{figure}[t]
\centering
\includegraphics[width=\columnwidth]{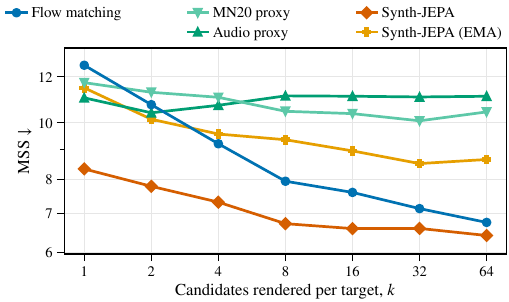}
\caption{Sound matching performance with inference-time best-of-$k$ selection of rendered audio.}
\label{fig:compute-secondary}
\end{figure}

Fig.~\ref{fig:compute-methods} highlights the difference between amortized inference and target-specific optimization.
Flow matching improves rapidly with additional ODE solver steps but then largely saturates beyond 16 steps.
Synth-JEPA, despite requiring more time to achieve flow matching's performance, continues to improve beyond this as the search budget increases.
Other search methods also benefit from greater budgets, but fail to match the performance of Synth-JEPA.
This indicates that additional test-time compute is more effectively exploited through target-specific search than through increasingly accurate amortized inference.

We can also scale test-time compute through an optional final best-of-$k$ selection of rendered candidates.
Specifically, we render a number of candidates through the synthesizer (meaning this approach is not strictly renderer-free) and select the one that most closely matches the target, according to the $L_1$ distance between log-Mel spectrograms.
The effects of this are illustrated in Fig.~\ref{fig:compute-secondary}.
Both flow matching and Synth-JEPA improve with best-of-$k$ search, but Synth-JEPA remains the strongest approach at all tested values of $k$.
Flow matching shows a steeper improvement over its $k=1$ baseline, suggesting that its conditional distribution contains high-quality solutions that are not reliably achieved with a single sample.

\subsection{Perceptual evaluation}

\begin{figure}[t]
\centering
\includegraphics[width=\columnwidth]{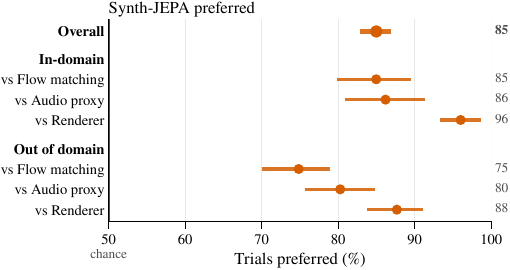}
\caption{Pairwise listening tests. Participants compared a reference to two sound matching attempts and selected the closest. Points show proportion of trials in which Synth-JEPA was preferred, with 95\% confidence intervals computed by bootstrap resampling of participants.}
\label{fig:listening}
\end{figure}

We conducted an online reference-conditioned pairwise listening test comparing Synth-JEPA with flow matching, log-Mel RITL search, and the log-Mel proxy.
Eighteen listeners judged 16 targets (8 in-domain, 4 NSynth, 4 FSD50K) across 3 baselines, giving 48 pairwise trials per participant.
Trial order and A/B assignment were randomized.
The stimuli in each trial were collectively normalized to a shared loudness of $-23$\,LUFS, thus preserving relative level differences
between the reference and reconstructions.
Listeners were asked to use headphones or high quality speakers.
Four attention-check trials, containing the ground truth as an option, were interleaved with the real trials.
All participants passed these checks.

Listeners preferred Synth-JEPA in 85.1\% of trials overall
(Fig.~\ref{fig:listening}), with preference remaining high both in-domain
(89.1\%) and out-of-domain (81.0\%).
Preference was strongest against RITL and weakest against flow matching.


\section{Conclusion}

We introduced Synth-JEPA, which learns mutually predictive audio and parameter representations from paired synthesizer data and uses them directly as a renderer-free objective for parameter search.
Our results show that it consistently outperforms parameter prediction baselines.
Comparisons within the renderer-in-the-loop and differentiable renderer-free settings indicate that Synth-JEPA's particular embedding space, learned directly from audio--parameter correspondences, is a more effective search objective than generic audio representations.
Accordingly, listeners preferred sound matches produced by Synth-JEPA in 85\% of trials in a pairwise listening test. 

While the model we present here was not trained for non-synthesizer audio, the generality of the JEPA framework permits the inclusion of various self-supervised learning objectives, thus allowing further methods for shaping the embedding geometry.
We plan to explore this in future work, as well as the extension of the framework to further synthesizers.

\newpage
{
\small
\section{Acknowledgements}
Haokun Tian is supported by the EPSRC UKRI Centre for Doctoral Training in Artificial Intelligence and Music (grant number EP/S022694/1).

\section{COMPLIANCE WITH ETHICAL STANDARDS}
\vspace{-1mm}
All participants provided informed consent for their participation. Data was collected in line with the principles of the Declaration of Helsinki.
\vspace{-2mm}
\bibliographystyle{IEEEbib}
\bibliography{references}
}

\end{document}